\documentclass[manuscript]{aastex}

\shorttitle{Homologous white light solar flares}
\shortauthors{Romano et al.}

\begin{document}


\title{Homologous white light solar flares driven by photosperic shear motions}


\author{P. Romano\altaffilmark{1}, A. Elmhamdi\altaffilmark{2}, M. Falco\altaffilmark{1}, P. Costa\altaffilmark{1}, A. S. Kordi\altaffilmark{2}, H. A. Al-Trabulsy\altaffilmark{2} and R.M. Al-Shammari\altaffilmark{2}}

\email{prom@oact.inaf.it}




\altaffiltext{1}{INAF - Osservatorio Astrofisico di Catania,
              Via S. Sofia 78, 95123 Catania, Italy.}
\altaffiltext{2}{Department of Physics and Astronomy, King Saud University,
							11451 Riyadh, Saudi Arabia}
										


\begin{abstract}
We describe the peculiarity of two recurrent white light flares occurred on Sept. 06, 2017, in the super Active Region (SAR) NOAA 12673, with a time interval, between their peaks, of about 3 hours. These events of X2.2 and X9.3 GOES class are very important not only for their high level of emission and for the visible effects on the lower layers of the solar atmosphere, discernible as clear white light ribbons, but as well for the strong horizontal photospheric motions which seemed to drive them. In fact, we observed for several hours before the flare occurrence the displacement of a negative umbra located in the main delta spot of the Active Region. We measured velocities up to 0.6 km s$^{-1}$. The strong and persistent shear motion of the photospheric structures together with the high intensity of the magnetic flux involved by these events can be considered responsible for the new energy continuously supplied into the magnetic system. From the timing of the emissions at different wavelengths we were able to provide some constraints for the modeling of such events.

\end{abstract}


\keywords{Sun: photosphere --- Sun: chromosphere --- sunspots --- Sun: surface magnetism}


\section{Introduction}

The secondary effects of high intensity flare are sometimes visible at photospheric level with the appearance of bright ribbons over the sunspot groups. This kind of events are called white light flares \citep[WLFs;][]{Sve66, Mac89, Fan95, Din03} considering they can be recognizable in the continuum of the solar spectra or in white light images. The occurrence of such events is infrequent as they usually require complex enough magnetic fields capable to trigger magnetic reconnection processes whose effects are also visible in the photosphere. Nevertheless, advances in high spatial resolution observations have provided more opportunities to detect WLFs even in less energetic events \citep{Jes08, Wan09}. This kind of flares are also reported in other stars, in young as well as in magnetically interacting stars, RS CVn type and magnetically active M dwarfs.
 
The origin of the continuum emission is still controversial. Initially, it is thought that the emission of WLFs is due to the particular high energy of some electron beams able to penetrate down into the photosphere, where they heat the photospheric plasma, and eventually radiates thermally \citep{Hud16}. However, recent simulations efforts argued that the electron beams are not able to reach the photosphere, but indeed their energy is already deposited at the chromospheric level \citep{Che10}. \citet{Fle08} have suggested that the Alv\'{e}n-wave pulses may be adequate to transport the energy from the reconnection site to the photosphere. Recently, another alternative interpretation of the WLFs has been further proposed, the so called $"$backwarming model$"$ \citep{Nei93, Din03, Met03, Che10}. In the cited model the photospheric emission is stimulated by the radiative heating from the chromosphere.

We report in this Letter the observations of two homologous WLFs occurred on Sept. 06, 2017, in AR NOAA 12673, just three hours away. Homologous flares are particular phenomena where the initial magnetic configuration, that was able to drive the event, is reformed after a previous one \citep{Nit01, DeV08, Arc14, Rom15}. If the WLFs consist rare events, homologous WLFs are even more rare and exclusive. We believe that the WLFs, subject of the present letter analysis, can provide some clues for a better understanding on how the WLFs are able to release their energy so down in the solar atmosphere and also on how homologous flares may occur in a short time intervals.

A description of the analyzed dataset is given in Sect. 2, together with an outline of the method for the photospheric horizontal velocity estimation. A depiction of the two events, at different layers of the solar atmosphere, is presented in Sect. 3. The main results are highlighted in Sect. 4. Finally we discuss and interpret our findings in Sect. 5.

\section{Data and analysis}

At the descending phase of the current Solar Cycle 24, on September 2017, the Super Active Region (SAR) NOAA 12673 has been observed on the Sun,  unleashing more than 40 C-, about 20 M- and 4 X-GOES class flares. Although beyond the scope of our present investigation, we mention here that SAR 12673 was subject of an unusual $sunquake$ phenomena caused by the X9.3 powerful flare\footnote{Details on this fact are reported by "HMI Science Nuggets" (http://hmi.stanford.edu/hminuggets/?p=2010) }.

In this Letter, we focused on the two X-GOES class flares occurred on Sept. 06 during a time interval of few hours (Figure \ref{fig1}). According to the reports of the National Geophysical Data Center (NGDC\footnote{ftp://ftp.ngdc.noaa.gov/}) the first event of class X2.2 started at 08:57 UT and reached its peak at 9:10 UT, while the second event of class X9.3, the most powerful since more than a decade, started during the main phase of the previous one, at 11:53 UT, and reached its peak at 12:02 UT.

For the description of the main aspects of these flares in the corona we used images taken at 171 \AA{} channel by the Atmospheric Imaging Assembly \citep[AIA;][]{Lem12} onboard the Solar Dynamic Observatory \citep[SDO;][]{Pes12}, with a time cadence of 12 s and a pixel size of about 0\farcs6. We considered the images taken on Sept. 06, from 8:00 UT to 20:00 UT.

We also adopted the Space-weather Active Region Patches (SHARPs) data \citep{Hoe14} acquired by HMI \citep{Sch12}. We analyzed continuum filtergrams and the vector magnetograms obtained in the \ion{Fe}{1} line at 617.3 nm by HMI on Sept. 06, from 00:00 UT to 23:59 UT, with a gap between 05:46 UT and 08:34 UT, having a time cadence of 12 min and a pixel size of 0\farcs51. In order to evaluate the horizontal velocity fields we applied the Differential Affine Velocity Estimator method for vector magnetograms \citep[DAVE4VM;][]{Sch08} adopting an apodization window with a full width at half maximum of 11 pixels (5\farcs5) to magnetograms taken with a time interval of 24 min.

To reconstruct the RHESSI signal for 12$-$25 keV and 25-50 KeV bands we exploit the data taken around the peaks of the two WLFs by means of the detectors 1$-$9, obtaining a spatial resolution of 4 arcsec. We applied the CLEAN method \citep{Hur02} to the data taken at 9:10 UT and 12:09 UT with an integration interval of 4 s. It is noteworthy here that different standard techniques (see \citet{Hur02}) produced similar results.

Moreover, images in the H$\alpha$ line at 8652.8 \AA{} taken from 8:50 UT to 9:20 UT and from 11:50 UT to 12:20 UT by the GONG network and by INAF-Catania Astrophysical Observatory have been used in our study.

\section{Description of the homologous flares}

\begin{figure}
    \includegraphics[trim=0 120 0 570, scale=.75]{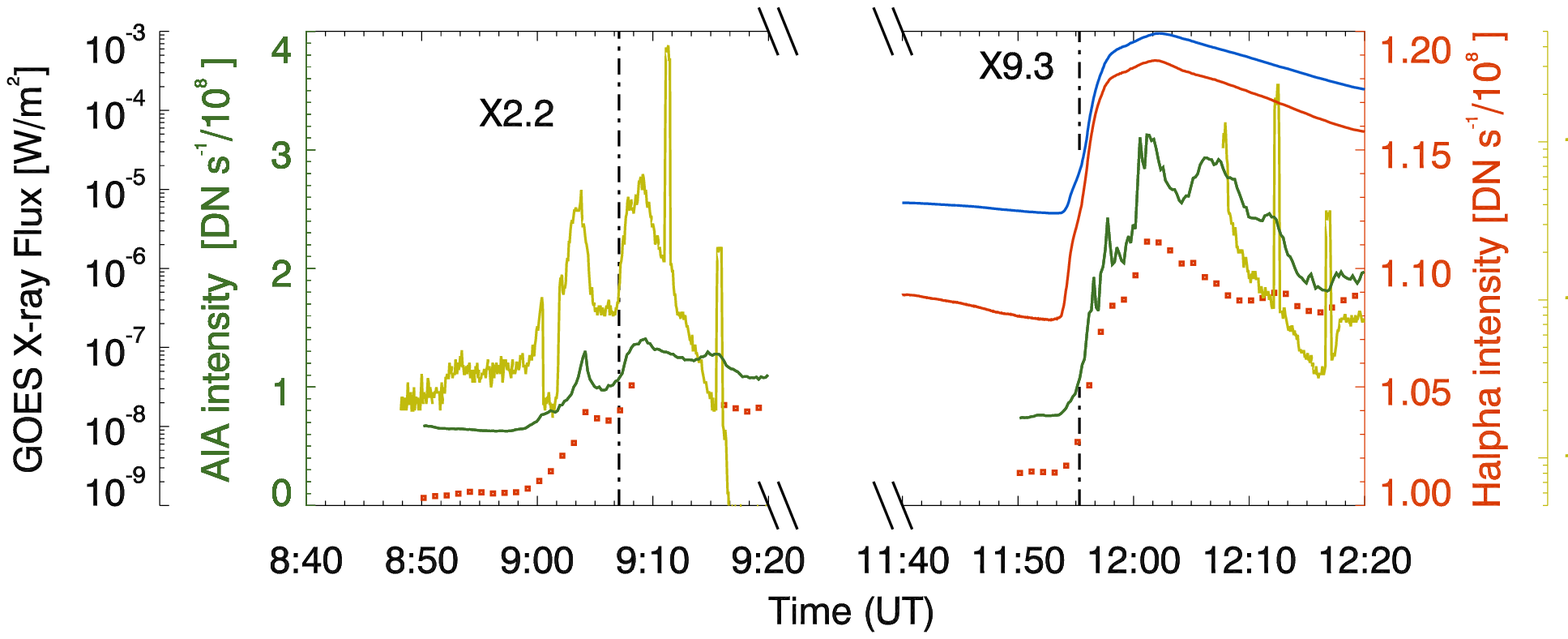}\\
    \includegraphics[trim=5 90 150 280, clip, scale=0.45]{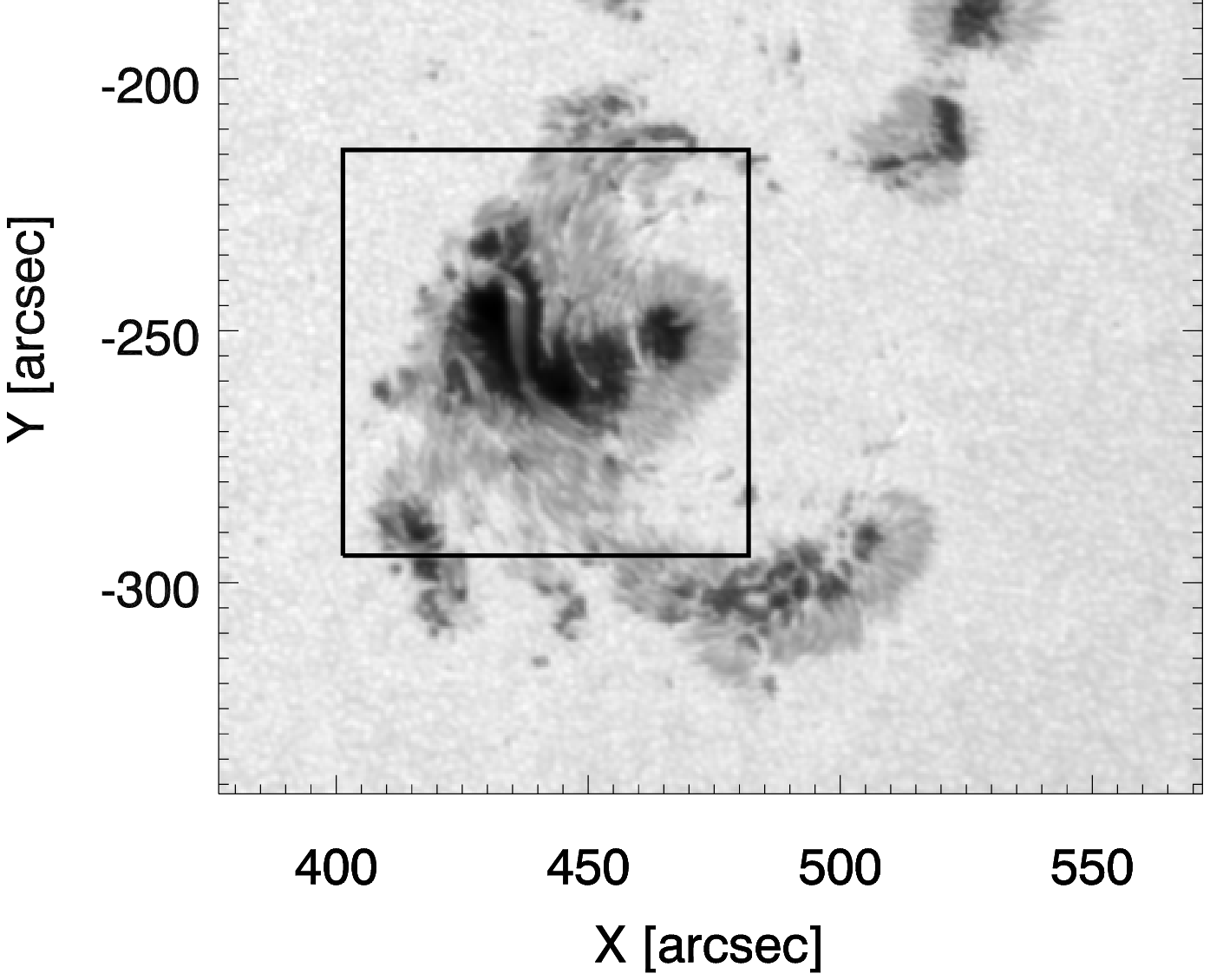}
    \includegraphics[trim=75 90 10 280, clip, scale=0.45]{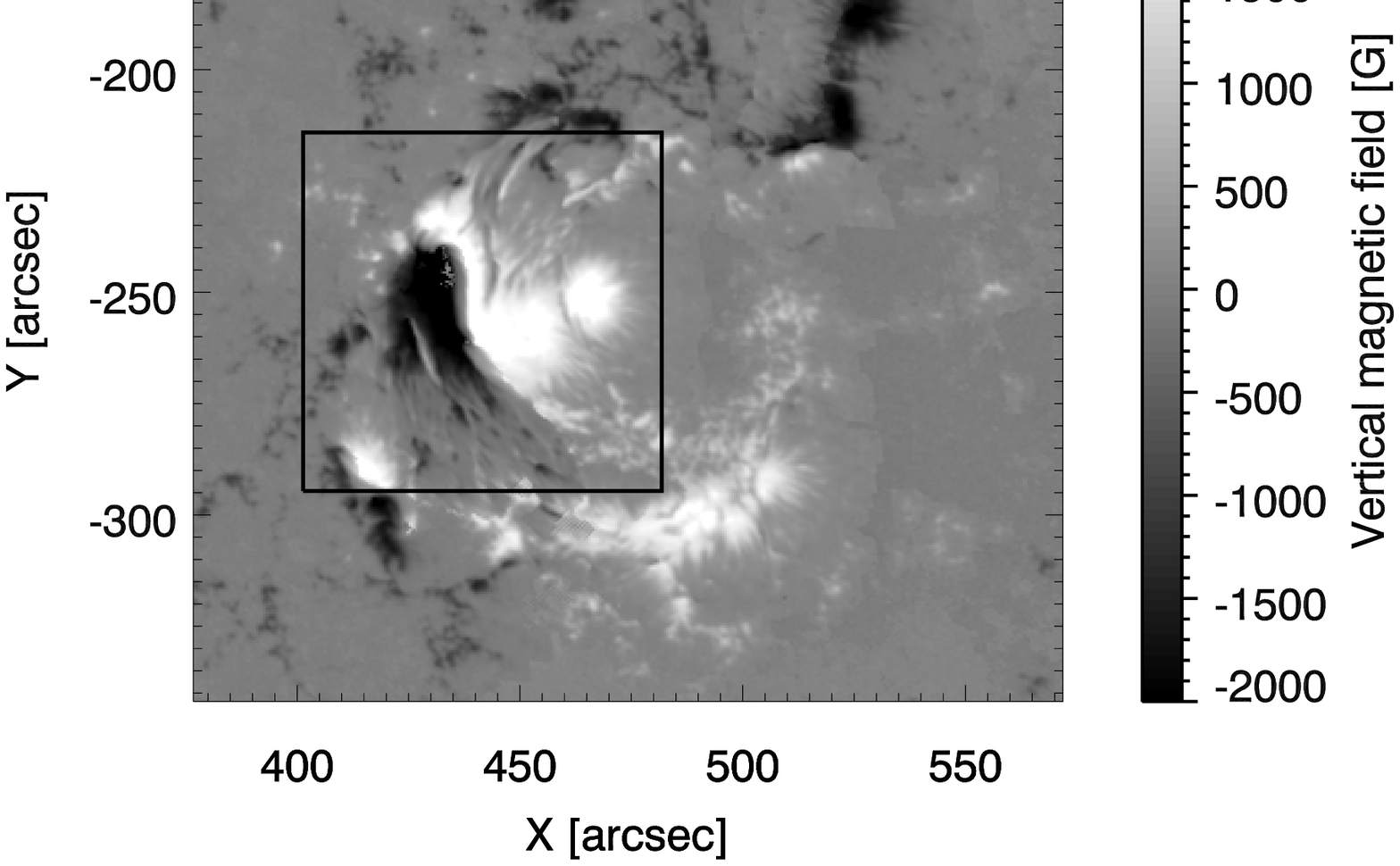}
  \caption{GOES X-ray light curves (blue curve: 1\,--\,8 \AA, red curve: 0.5\,--\,4 \AA). Overplotted are the HXR light curve deduced from RHESSI at 25$-$50 keV (yellow curve), SDO/AIA light curve taken at 171 \AA (green curve) and H$\alpha$ light curve (red boxes). The vertical dash-dotted lines indicate the first signatures of the WLFs in the photosphere. Bottom panels: SAR NOAA 12673 as seen in the continuum filtergrams (bottom-left panel) and in the magnetograms of the vertical component (bottom-right panel) taken by HMI/SDO at 617.3 nm. The boxes frame the portion of the field of view shown in Figures \ref{fig3}, \ref{fig4} and \ref{fig5}.\label{fig1}}
\end{figure}

On Sept. 06, 2017, the SAR NOAA 12673 (left bottom panel of Figure \ref{fig1}) appeared as composed by a main delta sunspot surrounded by a multiple of smaller sunspots. The delta spot was characterized by the presence of three umbrae inside the same penumbra: two of them located in the Western part were of positive polarity, while the Eastern one, with an elongated shape along the North-South direction, was of negative polarity (right bottom panel of Figure \ref{fig1}). The other sunspots, located in the North-West and South-West sides with respect to the main sunspot, were of negative and positive polarities, respectively.

\begin{figure*}
    \includegraphics[trim=5 155 150 330, clip, scale=0.42]{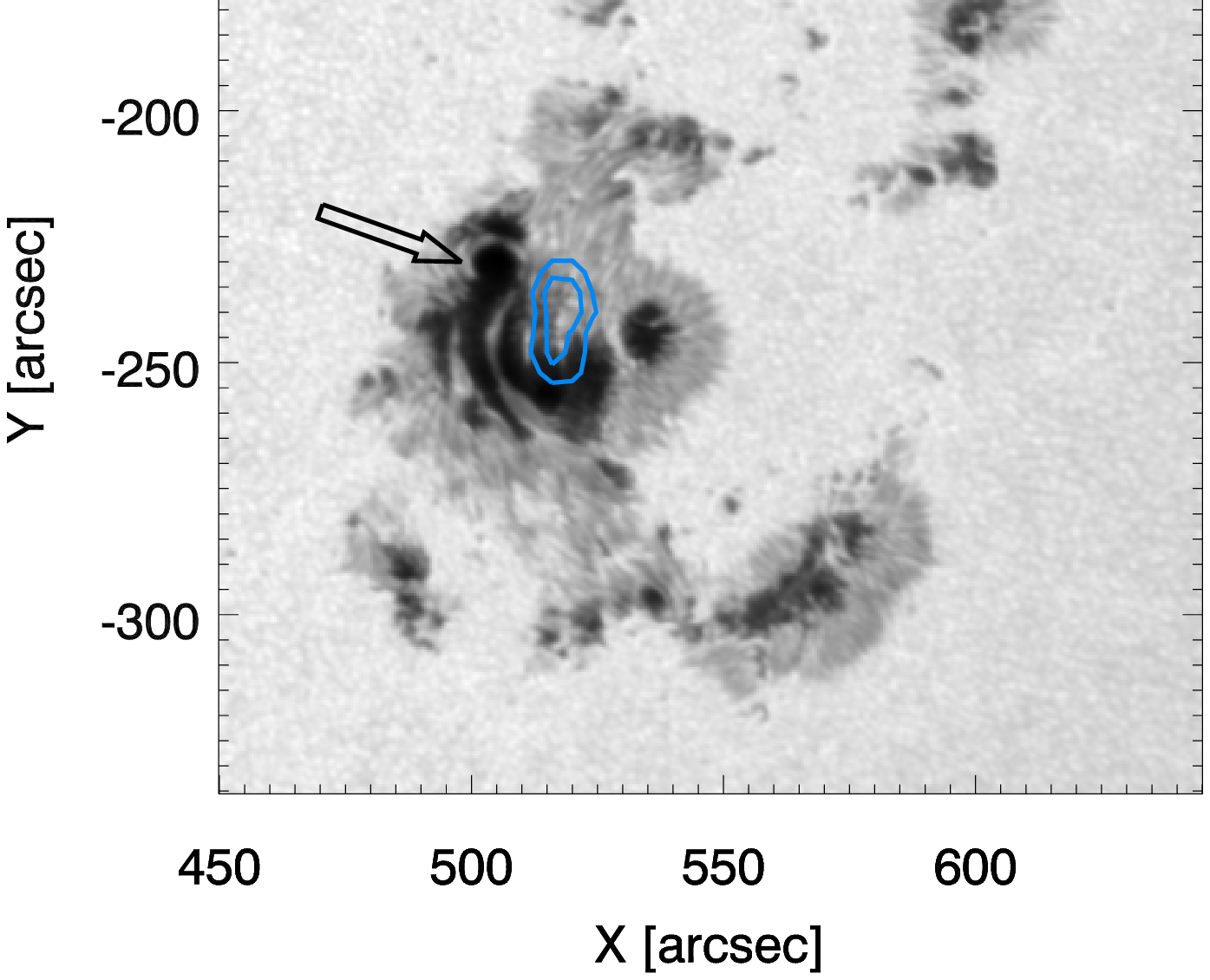}
    \includegraphics[trim=75 155 10 330, clip, scale=0.42]{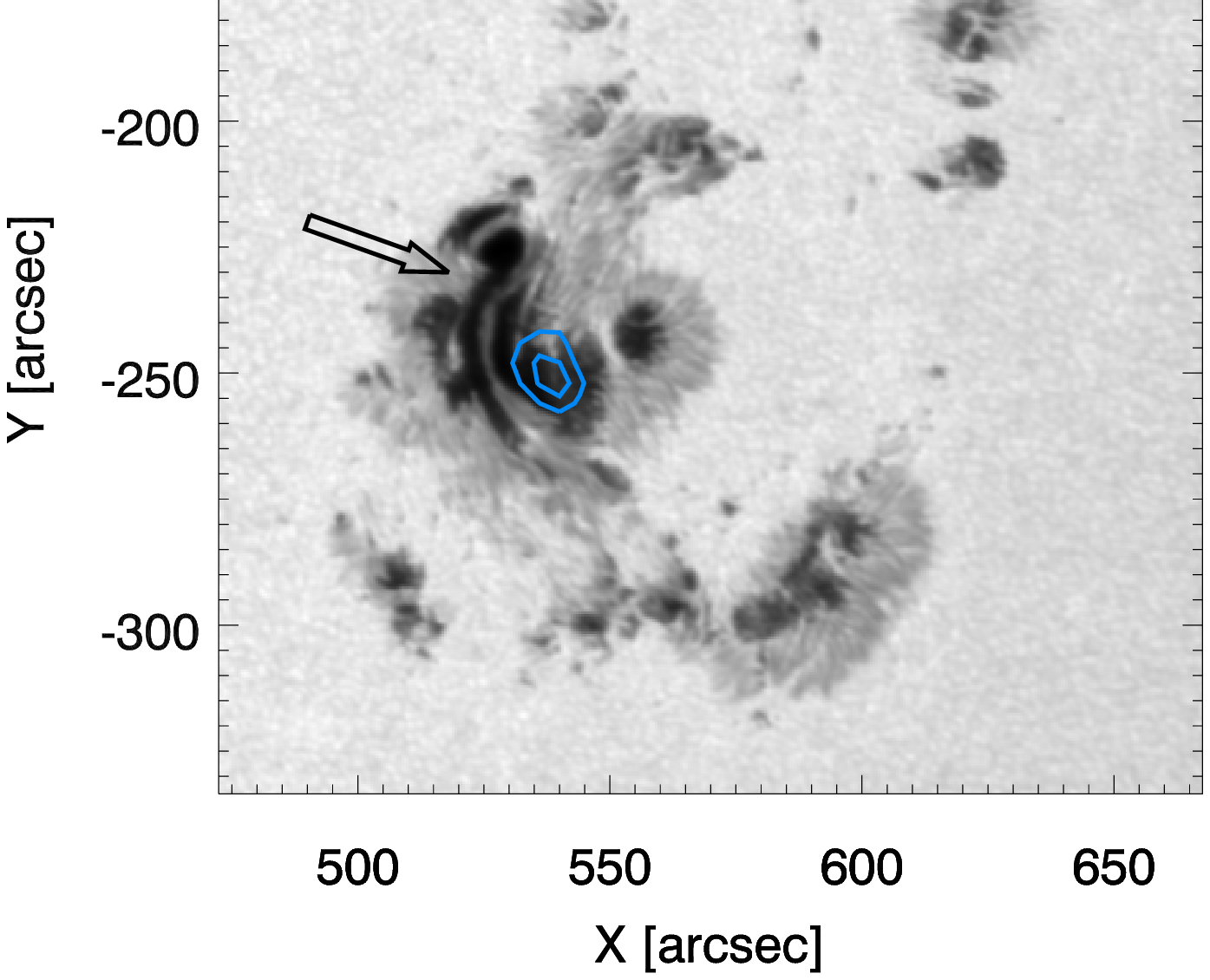}\\
		\includegraphics[trim=5 155 150 330, clip, scale=0.42]{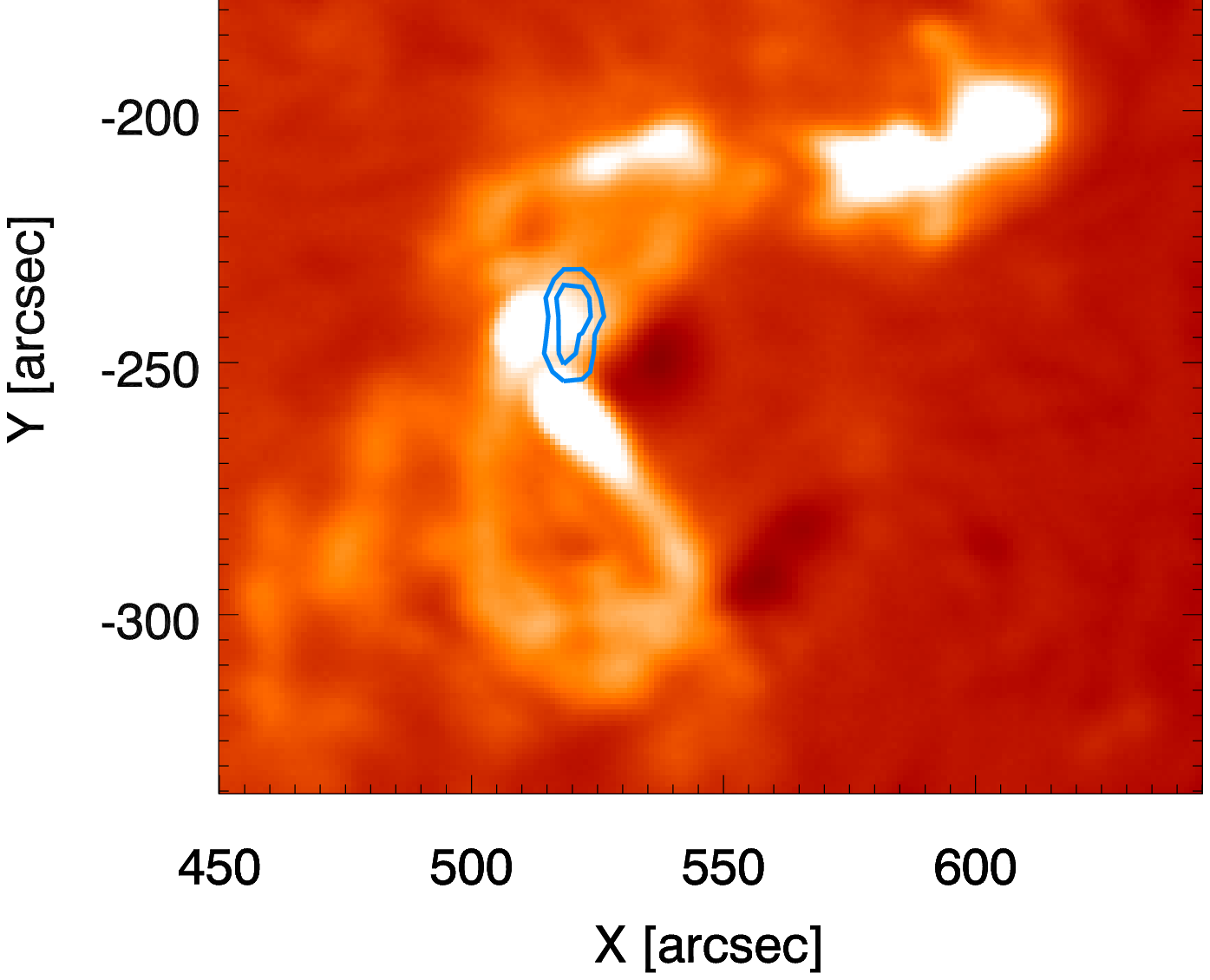}
    \includegraphics[trim=75 155 10 330, clip, scale=0.42]{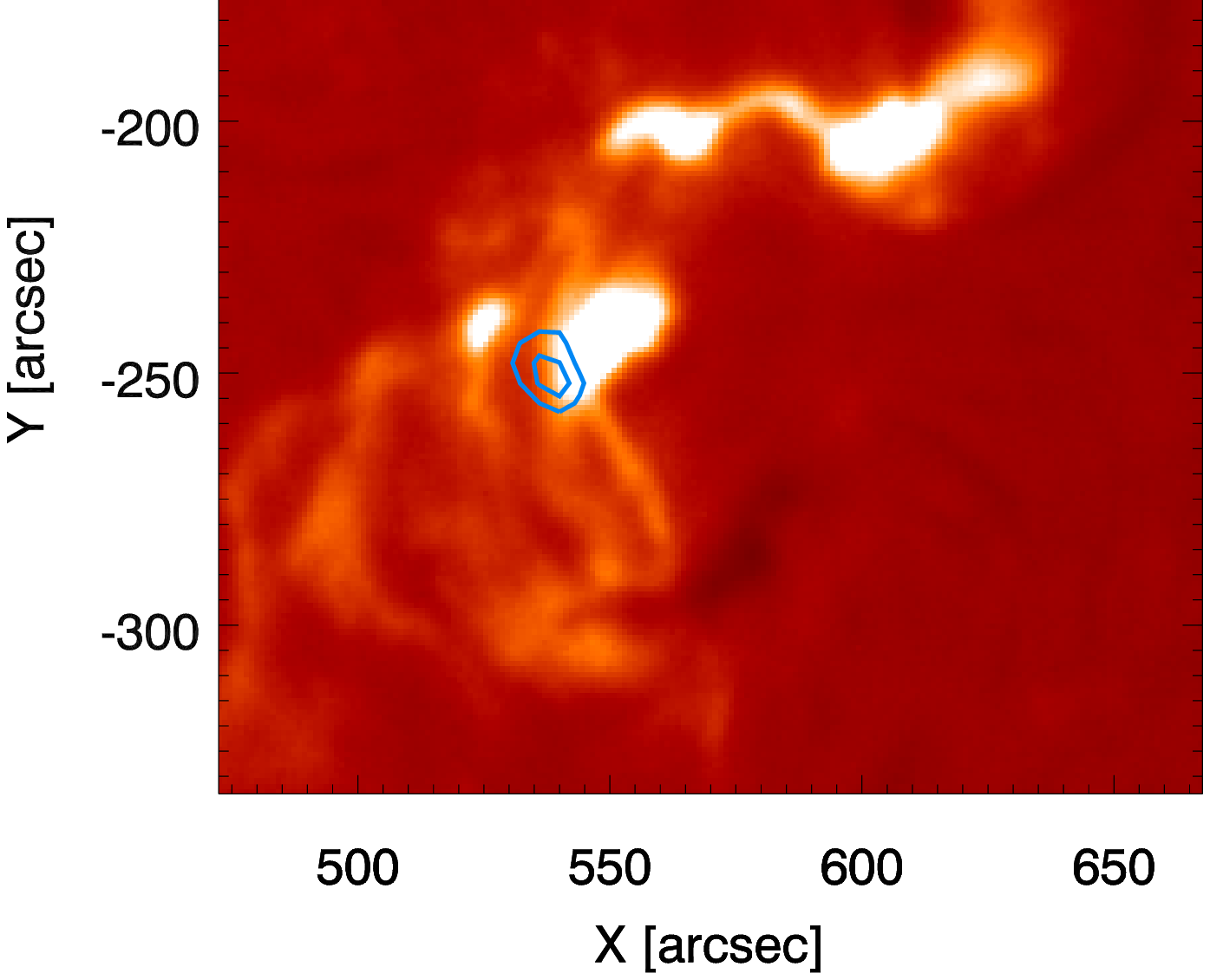}\\
		\includegraphics[trim=5 110 150 330, clip, scale=0.42]{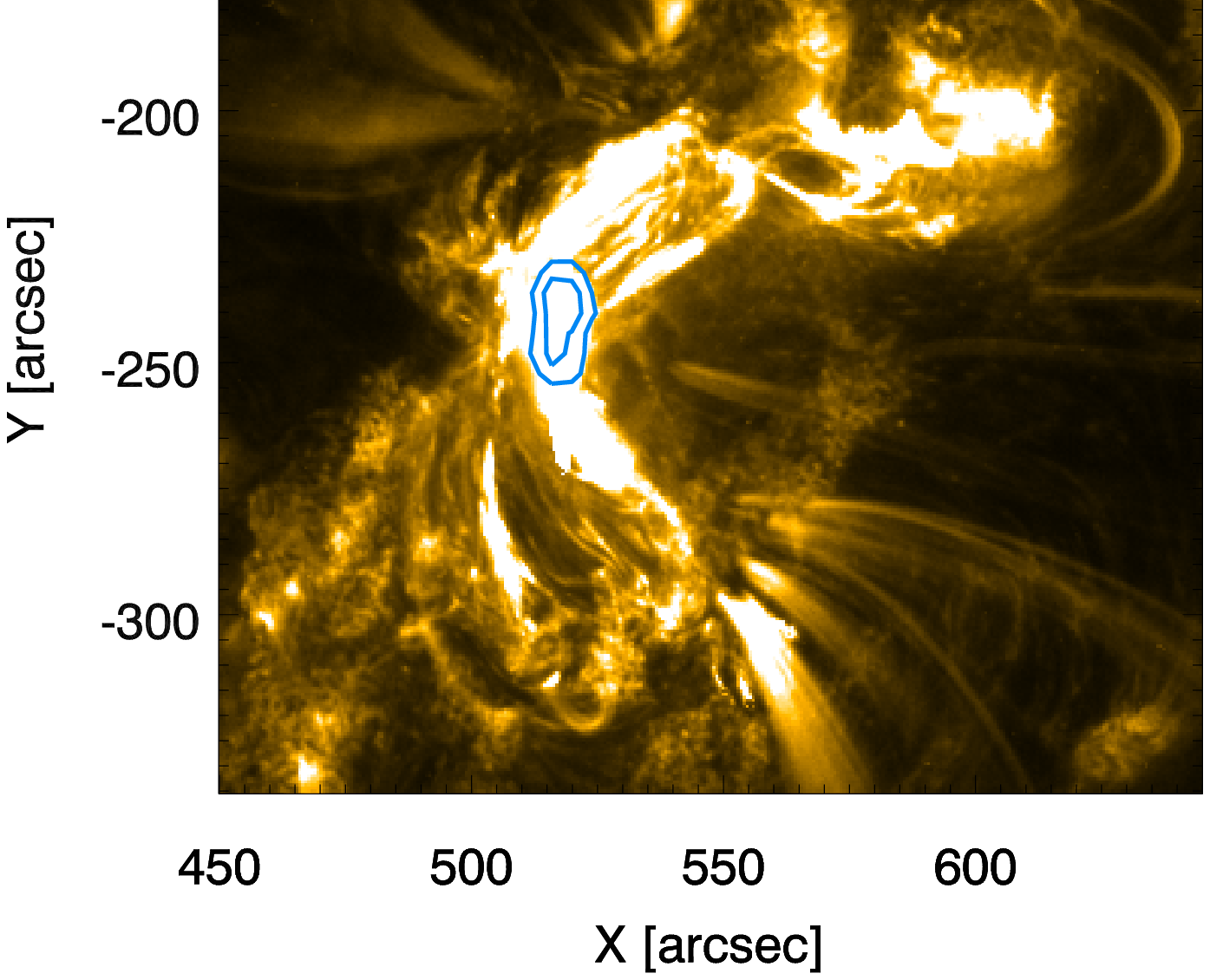}
    \includegraphics[trim=75 110 10 330, clip, scale=0.42]{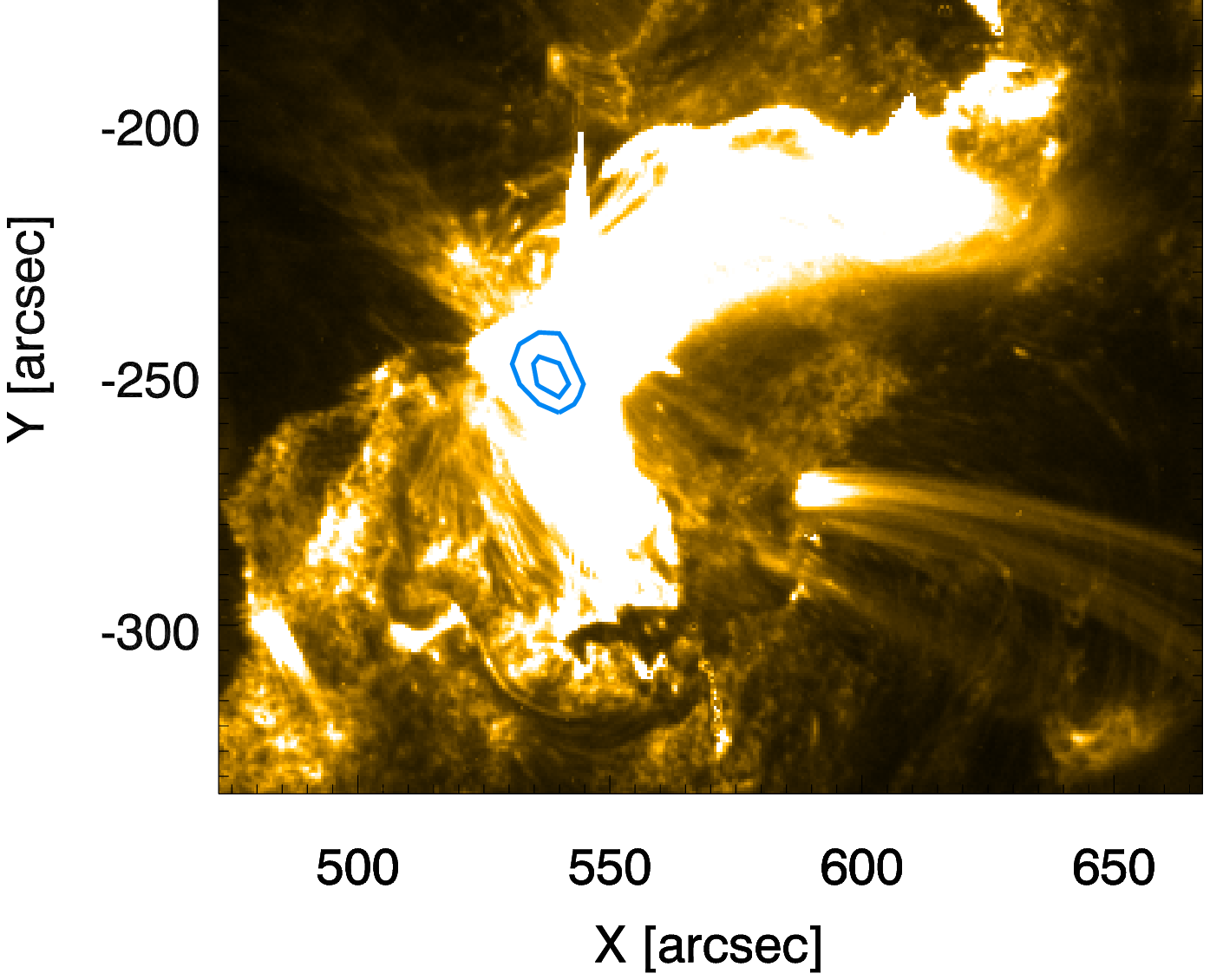}\\
  \caption{AR NOAA 12673 as seen during the main phase of the X2.2 (left column) and the X9.3 (right column) GOES class flares: in the first raw the HMI continuum filtergrams, in the second raw the H$\alpha$ images taken by INAF-Catania Astrophysical Observatory (left) and by the Cerro Tololo station of the GONG H$\alpha$ network (right) and in the third raw the AIA images at 171 \AA. The overlaid contours indicate 70\% and 90\% of the flux peak at 12$-$25 keV taken by RHESSI around the peaks of the WLFs. The arrows in the first raw indicate the elongation of the umbra.\label{fig2}}
\end{figure*}

The two monster X-class flares involved more or less the whole SAR, as seen in the images taken during the main phases at photospheric, chromospheric and coronal levels (Figure \ref{fig2}). The images acquired by INAF - Catania Astrophysical Observatory in the core of the H$\alpha$ line (middle left panel of Figure \ref{fig2}) evidence where most of the chromospheric emission was located during the first flare of X2.2 GOES class. We can identify three regions where the detector reached its saturation level: one above the delta spot, one at the northern edge of the penumbra and one above the smaller negative sunspots in North-West side relative to the delta spot. The rough $S$-shape of the chromospheric ribbons is also visible in corona at 171 \AA{} channel (bottom left panel of Figure \ref{fig2}). The second flare, stronger than the previous one, exhibits a noticeable increase of the same ribbons accompanied with an extension of their length at chromospheric level (middle right panel of Figure \ref{fig2}). The exceptional intensity of the X9.3 GOES class flare is also documented by the saturation of the AIA detector which remained visible after about 30 min from the peak of the event at 12:30 UT (bottom right panel of Figure \ref{fig2}). The fact that the two flares occurred in a time interval of about 3 hours and that the location of the ribbons was roughly the same allowed us to confidently catalog them as recurrent and homologous flares.

The comparison between the HMI continuum filtergrams, obtained at 9:34 UT and 12:34 UT, emphasizes a clear elongation of the negative umbra of the delta spot (see the arrows in the top panels of Figure \ref{fig2}). This required a deep investigation in order to understand whether it is the result of a photospheric magnetic reconfiguration or due to the effect of a strong shear. 

The contours of 70\% and 90\% of the flux peak at 12$-$25 keV, obtained from RHESSI observations at 09:10 UT and at 12:09 UT (see the contours in Figure \ref{fig2}), indicate that the hard X-ray (HXR) sources for both WLFs do not seem to be located above the elongated negative umbra, but rather at its western side, i.e. in the center of the delta spot. Although we cannot exclude that the location of the X-ray source may be affected by some projection error due to the high longitude of the SAR, we can be confident that in both events the heating sources involve the same magnetic system of the SAR. Additionally, we also note that the contours are located above the southern ribbon in the H$\alpha$ images and in the center of the S-shaped structures visible at 171 \AA.

\begin{figure*}
		\includegraphics[trim=0 100 150 300, clip, scale=0.4]{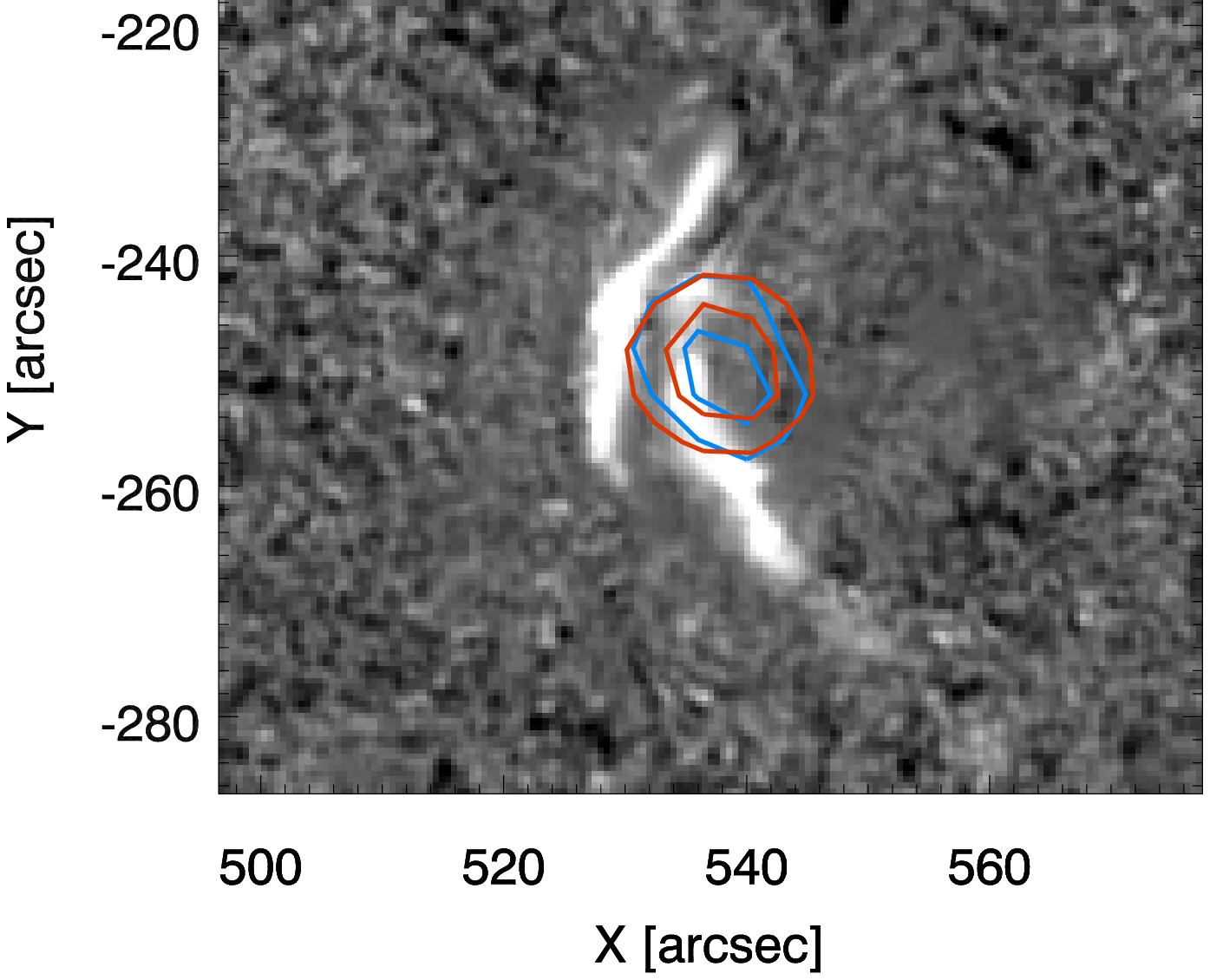}
    \includegraphics[trim=73 100 0 300, clip, scale=0.4]{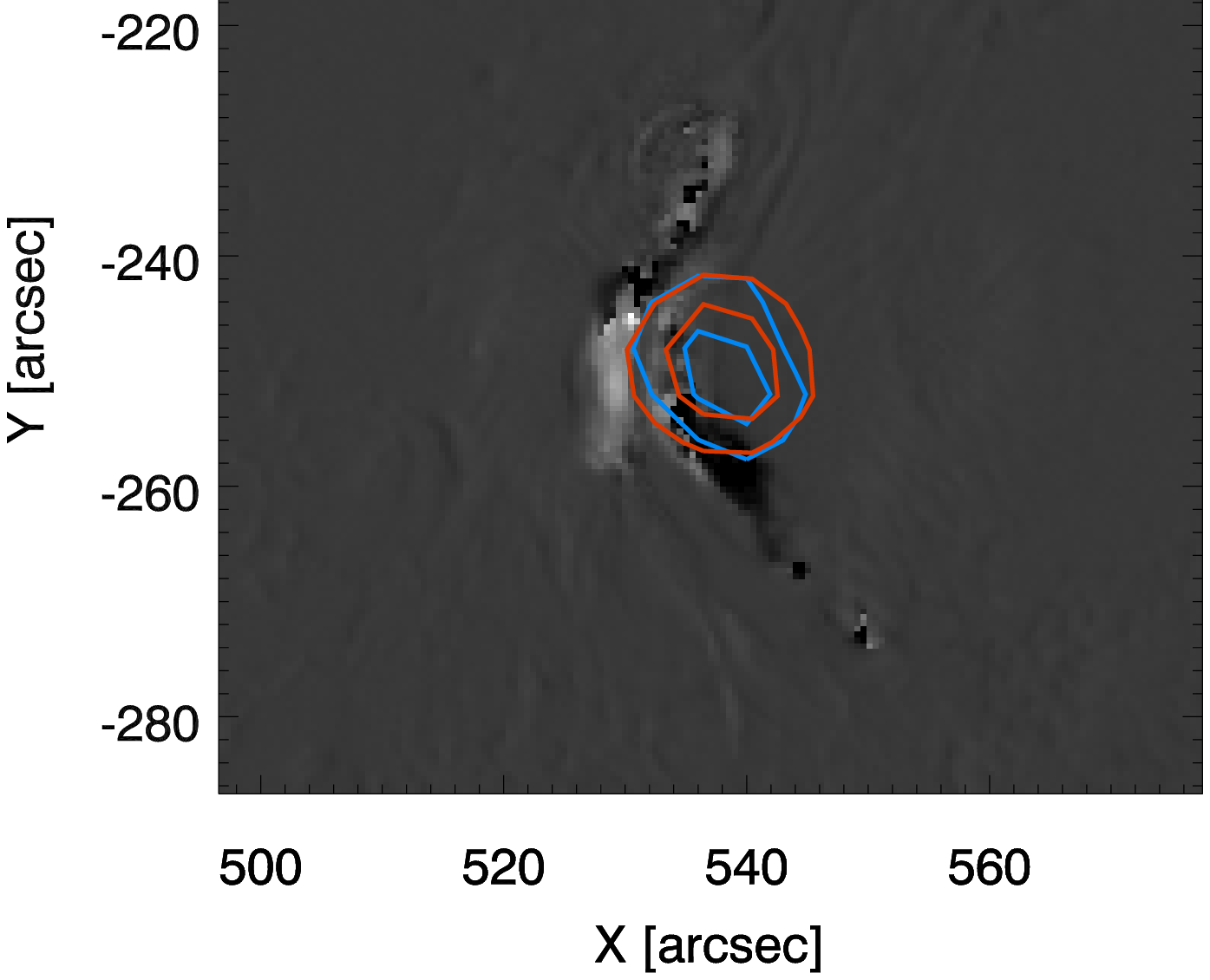}\\
		\includegraphics[trim=0 100 150 300, clip, scale=0.4]{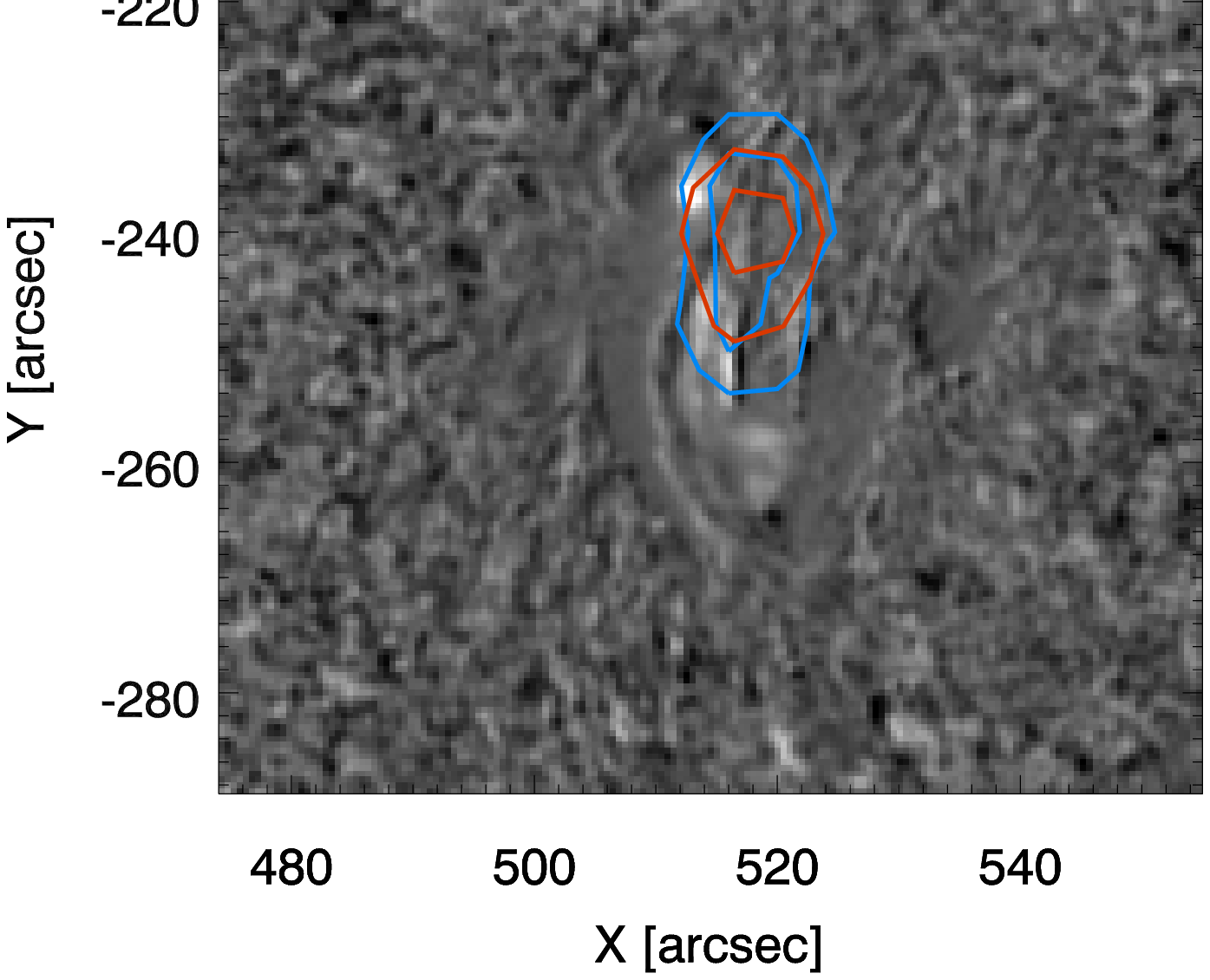}
    \includegraphics[trim=73 100 0 300, clip, scale=0.4]{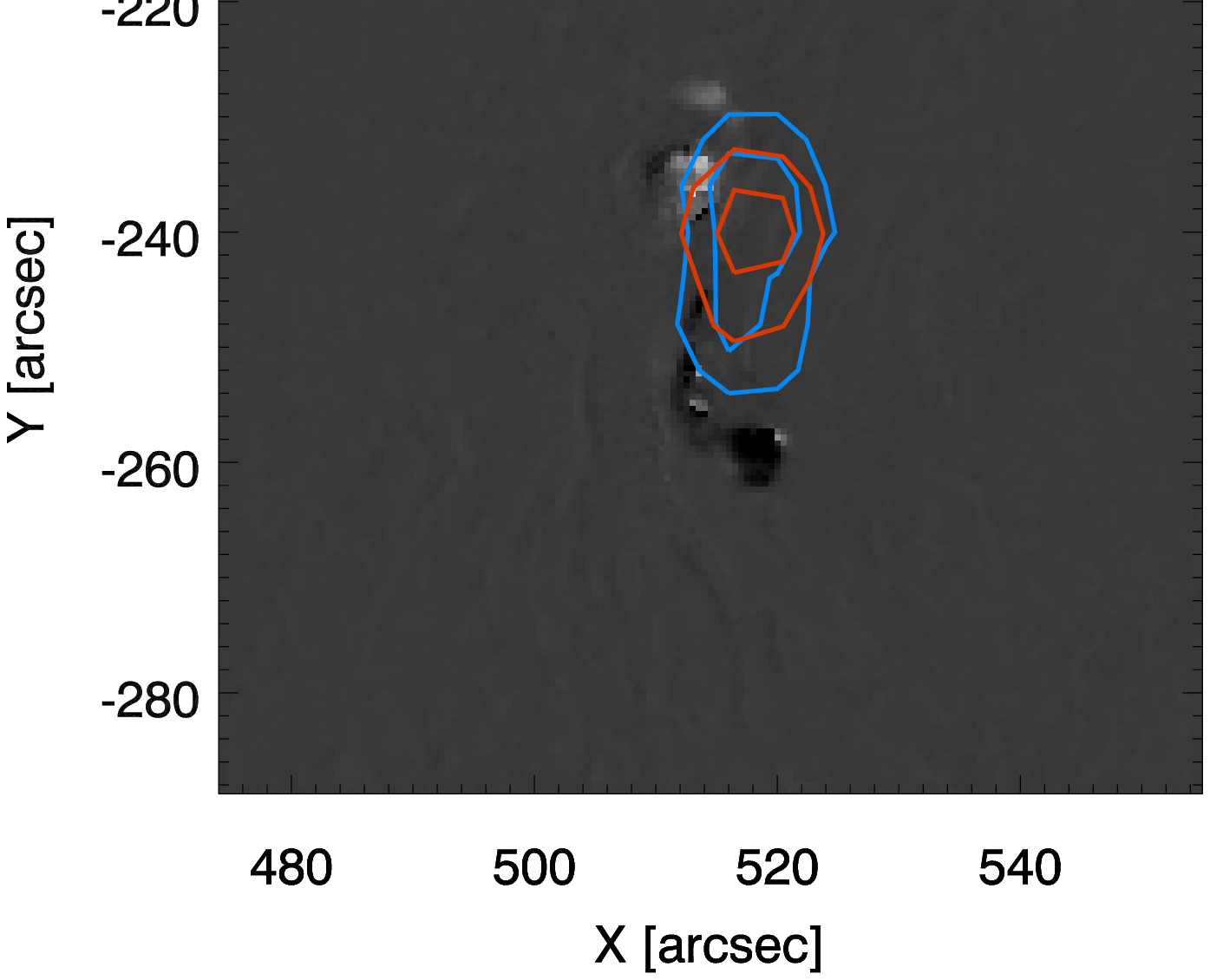}\\
  \caption{Running-difference images during the peak intensity of the WLFs associated ribbons, X9.3 (top panels) and X2.2 (bottom panels).  The blue and red overlaid contours indicate 70\% and 90\% of the flux peak registered by RHESSI around the peaks of the WLFs at 12$-$25 keV and 25$-$50 keV, respectively. More details about the WLFs phase, and generally about AR 12673 evolution during Sept. 06, can be seen in the accompanying online movie ($WLFs\_AR12673.wmv$; see text)\label{fig3}}

  \end{figure*}

To highlight the presence and nature of the WLFs associated ribbons, we performed a running difference method in the HMI continuum filtergrams as well as for magnetograms, adopting a 2 min time interval between successive images. Figure \ref{fig3} reports the resulted processed image around the flare events peaks. For the X9.3 case, we clearly recognize an intensity enhancement and the residual white light ribbons in both continuum and magnetogram images, (top panels of Figure \ref{fig3}), with a semi-circular shape of latitudinal extended length as high as 40\farcs. The earliest WL ribbons sign appeared during the impulsive phase at 11:55 UT, then the ribbons rapidly separate and disappear around 12:09 UT. The HXR contours at 12$-$25 keV (blue line) and at 25$-$50 keV (red line) are located above the Western white light ribbon. Furthermore, we also distinguish a WLF signature in the earliest X2.2 flare, depicted in the bottom panels of Figure \ref{fig3}, along the PIL, although the indicated ribbons appear less elongated compared to the severe X9.3 related ones. Here again, the X-ray sources are located in the Western side of the ribbon area. The evolution of the SAR on Sept. 06, and specifically of the associated WLFs can be seen in the accompanying online material ($WLFs\_AR12673.wmv$). The movie utilizes a 2 min cadence images (running-difference HMI continuum, magnetograms, inversed colors AIA 171 \AA{} channel and GONG H$\alpha$) where the differential rotation effects have been removed.

The comparison among the light curves obtained by GOES, RHESSI, AIA/SDO and GONG, averaging the intensity in the field of view indicated by the boxes in the bottom panels of Figure \ref{fig1}, evidences a timing difference in the corresponding emissions at different wavelengths (in both WLFs; Figure \ref{fig1}). On the one hand, although there is a lack in the acquisition of GOES and RHESSI data during the flash phase of the first and the second event, respectively, we can distinguish a first intensity enhancement reported by the two GOES channels (see the red and blue lines in the top panel of Figure \ref{fig1}), then the beginning of the events in the HXR data (yellow curve), followed by that at 171 \AA{} (green curve) and lastly in the chromosphere (red squares). On the other hand, the peaks of these curves are not in the same order. In fact the H$\alpha$ peaks precede the peaks registered in the corona with the other instruments by few minutes. We also note multiple peaks at 171 \AA{} and in the HXRs, probably due to the occurrence of several reconnection events. The first signatures of the WLFs in the photosphere (vertical dash dotted lines in Figure \ref{fig1}) have been observed at 09:08:04 UT and 11:55:19 UT for the X2.2 and the X9.3 flares, respectively, i.e., after the rise phase and before the peak of the flares in the H$\alpha$ line.

\section{Horizontal velocity fields}

\begin{figure}
   \includegraphics[trim=5 390 10 10, clip, scale=0.85]{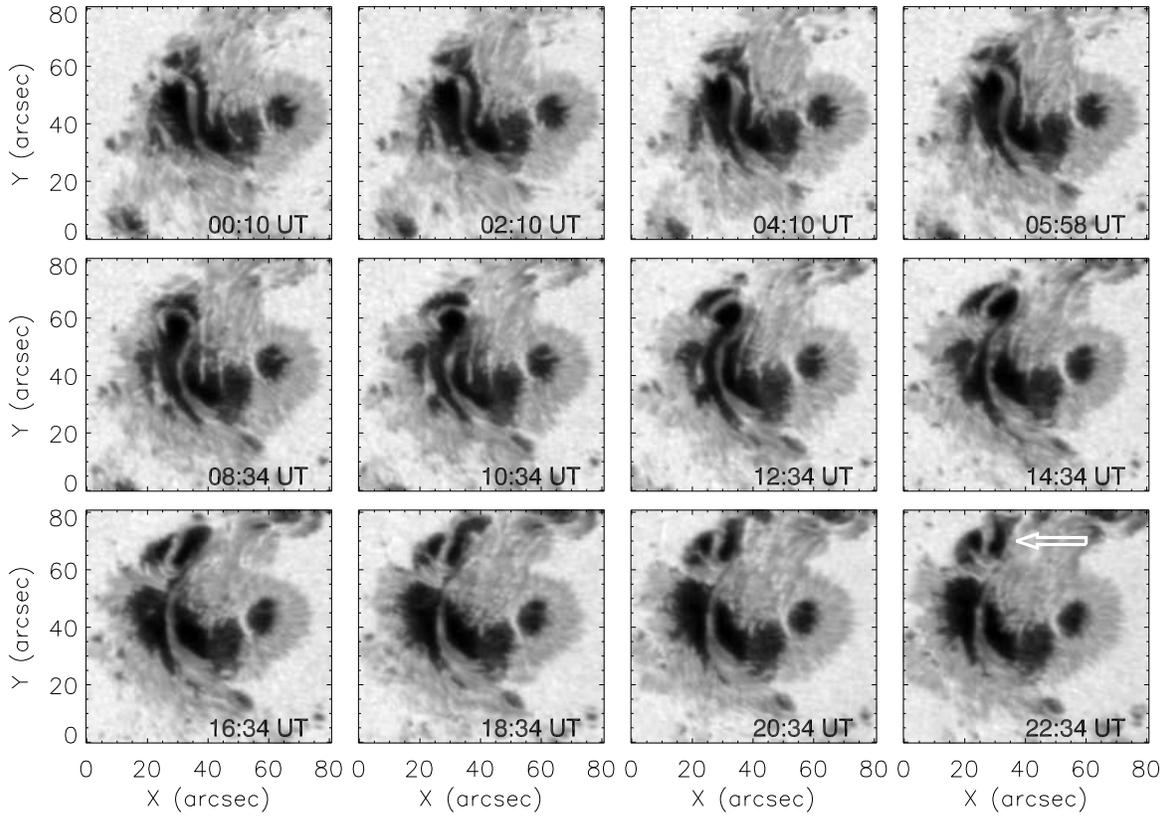}
  \caption{SAR NOAA 12673 as seen in the continuum filtergrams taken by HMI/SDO at 617.3 nm.\label{fig4}}
\end{figure}

Figure \ref{fig4} reports the temporal evolution of the main delta spot, on Sept. 06, at the HMI continuum filtergrams with a time cadence of 2 hours taking into account the gap in the SHARP dataset between 05:58 UT and 08:34 UT. Interestingly, we observe a clear stretching behavior of the Eastern umbra occurring till 18:34 UT, i.e. during the hours preceding the two flares and during the main phase of the second flare. Only when the umbra is splitted into two portions and a wide light bridge appears in the Northern part of the spot, the stretching stops and the delta spot seems to gain a phase of stability.

\begin{figure*}
    \includegraphics[trim=15 75 320 110, clip, scale=0.47]{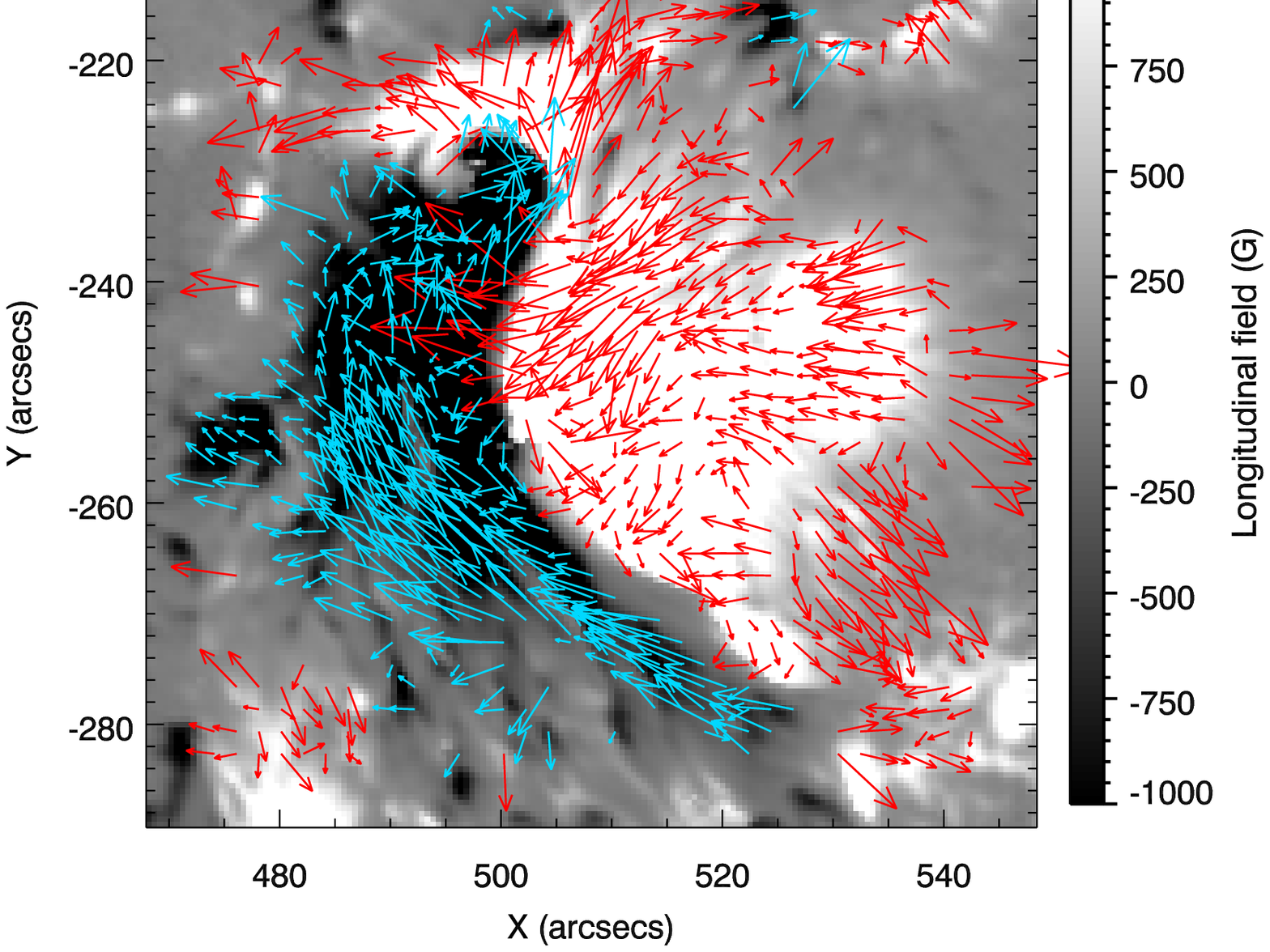}
    \includegraphics[trim=15 75 320 110, clip, scale=0.47]{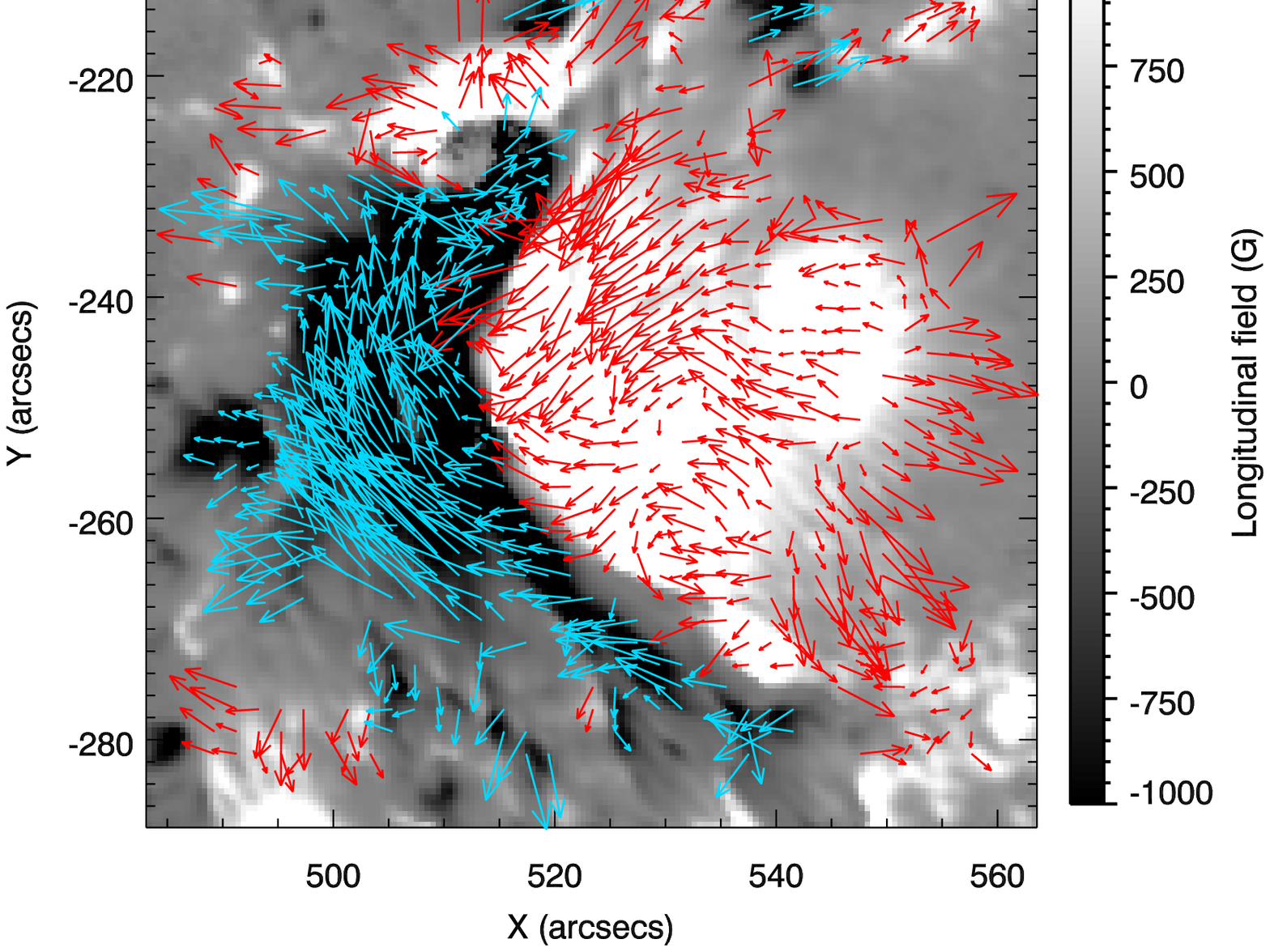}\\
		\includegraphics[trim=15 75 320 110, clip, scale=0.47]{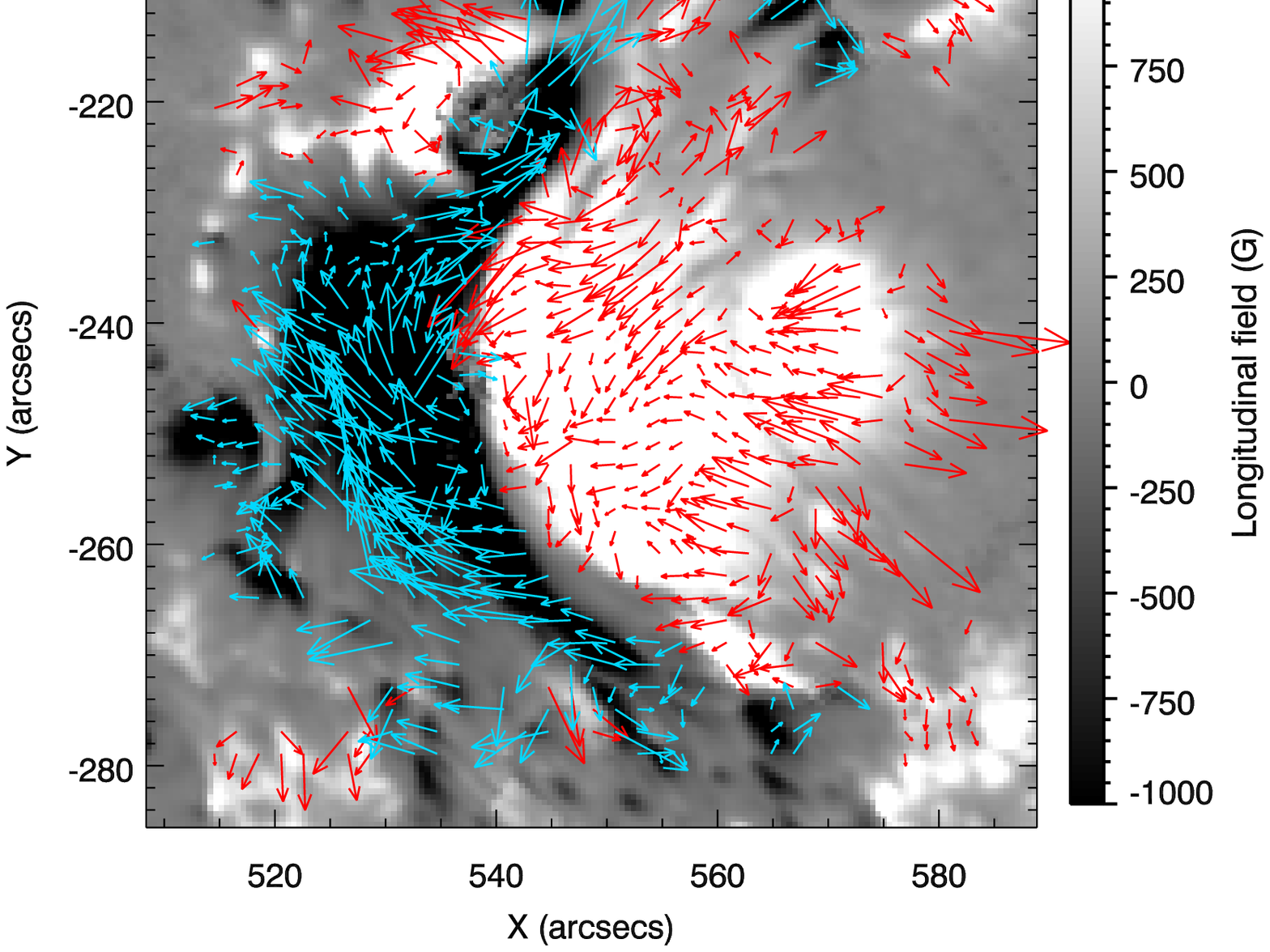}
    \includegraphics[trim=15 75 320 110, clip, scale=0.47]{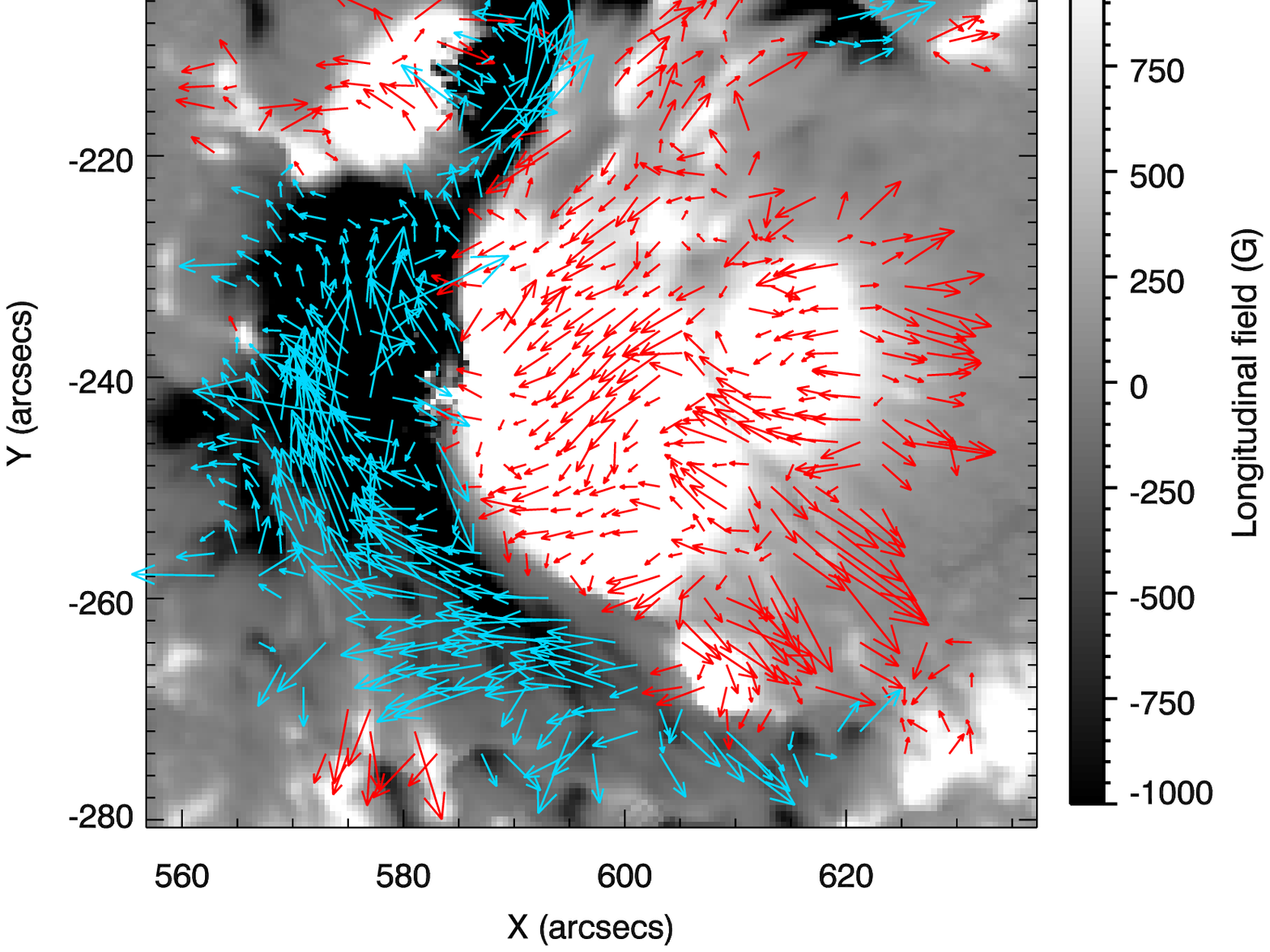}\\
  \caption{Horizontal velocity fields measured by means of the DAVE tracking processing method. The background images correspond to the average of the vertical component of the magnetic field measured by HMI during the time interval considered for the velocity computations. The red arrow in the top left corner indicates a scale velocity of 0.5 km s$^{-1}$. \label{fig5}}
\end{figure*}

To adequately quantify the velocity of the observed displacement, we apply of the DAVE tracking method to the vector magnetograms. We found that the negative umbra of the delta spot migrated northward with an average velocity of 0.4 km s$^{-1}$, reaching values as high as 0.6 km s$^{1}$. These peculiar motions persist during the whole day, although the highest velocities have been particularly registered before the flares onsets (top panels of Figure \ref{fig5}). A velocity decrease is indeed discernible already during the main phase of the X9.3 class flare (bottom left panel of Figure \ref{fig5}). As shown in the bottom right panel of Figure \ref{fig5}, the velocities of the magnetic features forming the main part of the negative umbra spread out in various directions after the flares. These velocity maps allow us also to notice the motion of the main positive polarity in opposite direction to the negative umbra, although with lower velocities. This apparent complex coupling of motions suggests the delta spot being a theater of a strong shear during the phases preceding the two flares and, in particular, during the time interval between their peaks.

\section{Discussion and conclusions}

The various unusual observational facts associated to the two homologous WLFs reported in this Letter allow us to draw some interesting conclusions about the possible physical mechanisms at the base of such events. We observed a fast and continuous displacement of a negative feature located inside a positive delta spot. We measured an average velocity of about 0.4 km s$^{-1}$ for several hours, not only before the beginning of first event but also during the time interval between the peaks of the two event. We excluded that the elongation of the negative polarity was an instrumental effect due to the inversion of the \ion{Fe}{1} line at 617.3 nm as a consequence of the strong flare emission, because we observed this effect only in the maps of the vertical component of the magnetic field taken by HMI during the time interval between 11:55 UT and 12:02 UT, i.e. during the peak of the second flare. In those magnetograms a portion of the negative polarity became apparently positive and a small negative patch appeared in the surrounding positive part of the sunspot. Moreover, the northward elongation of the negative polarity was confirmed by the evident stretching of the eastern umbra inside the penumbra of the delta spot. This northward displacement of the penumbra was also accompanied by the southward motions of the magnetic features forming the remaining part of the sunspot. Therefore, shear motions have been observed till the end of the second flare. Apparently, a similar photospheric configuration characterizes the delta spot before and after the two WLFs (compare the top left and bottom right panels of Figure \ref{fig4}). However, looking carefully at the development of the continuum frames taken between 8:34 UT and 16:34 UT (see all panels of Figure \ref{fig4}) we can notice that the small umbra indicated by the arrow at 22:34 UT is not the same umbra located at [35\farcs, 65\farcs] in the top left panel, but it is indeed the remaining portion of the elongated umbra, which separated between 14:34 UT and 16:34 UT.

The shearing motions provide an observational evidence of the existence, at photospheric level, of a possible mechanism capable to supply new energy into the magnetic system for many hours as well as after a first episode of energy release. In fact, according to \citet{DeV08}, recurrent flares can occur also when the coronal magnetic field is continuously sheared by photospheric motions. However, it remains unclear the cause of these shear motions and why we registered a variation in the proper motions of the SAR after the release of energy of the second event. We could argue that the shear motions persist after the first event, despite of the first huge release of free energy, because of a stable configuration of the magnetic field was not reached yet. Rather, the unchanged variation of the magnetic field configuration after the first flare was confirmed by the homology of the flares. The similar shape of the ribbons of the two flares at chromospheric level and the similar brightening in the corona corroborate the above mentioned scenario.

Although the determination of the coronal magnetic configuration responsible for the two WLFs is far from the main scope of the present Letter, it is noteworthy that the observed shearing motions can be located in a photospheric area corresponding to the central portion of the $S$-shape ribbons at chromospheric level and under the bundle of loops forming a sort of sigmoid at 171 \AA{} (see the accompanying online movie). Taking into account that the first brightening at coronal level in both flares occurred above the delta spot, where we observed the umbra elongation, we argue that the evolution at photospheric level may play a crucial role in the determination of the SAR instability. However, we plan to determine the topology at the base of the two WLFs and to verify the occurrence of some magnetic flux cancellations along the polarity inversion line in a forthcoming Paper.

The timings of emission in the different layers of the solar atmosphere indicate the continuum emission being observed some minutes after the first appearance of the ribbons at chromospheric level in the first event and, certainly, before the peaks of the H$\alpha$ light curves in both flares. This behaviour does not allow us to discriminate among the different mechanisms proposed to explain the continuum emission of the WLFs, but provides some useful constraints for the modeling of this phenomenon. In fact, we found for both events that the emission at photospheric level may be observed several minutes before the emissions in X-rays, EUV and H$\alpha$ lines (i.e., in the upper layers of the solar atmosphere) reached their maximum.

It is undoubtedly clear that there are still many un-answered questions and un-assessed key ingredients regarding the different mechanisms responsible of the reported peculiarities of these interesting events. Therefore, we believe that enlarging similar studies of this SAR-like class will help us to shed light on the mechanisms able to supply the energy during recurrent flares, on the mechanisms at the base of the photospheric signatures of the white light flares and as well on the complex dynamics of delta sunspots.

\acknowledgments

This work utilizes data obtained by INAF - Catania Astrophysical Observatory and by the Global Oscillation Network Group (GONG) Program, managed by the National Solar Observatory, which is operated by AURA, Inc. under a cooperative agreement with the National Science Foundation. The research by A. Elmhamdi was supported by King Saud University, Deanship of Scientific Research, College of Science Research Center.



{\it Facilities:} \facility{SDO (HMI, AIA), RHESSI, OACT:0.15m}.



\clearpage

\end{document}